\documentstyle[12pt]{article}
\begin{document}
\def\tfrac#1#2{{\textstyle{#1\over #2}}}
\def\half{\tfrac{1}{2}}
\def\bch{Q}
\def\Qop{{\bf Q}}
\def\Vop{{\bf V}}
\def\Uop{{\bf U}}
\def\p{\partial\phi_2}
\def\pp{\partial^2\phi_2}
\def\ppp{\partial^3\phi_2}
\def\bph{{\bf \varphi}}
\def\brh{{\bf \rho}}
\def\be{\begin{equation}}
\def\ee{\end{equation}}
\def\ba{\begin{eqnarray}}
\def\ea{\end{eqnarray}}
\def\dif{\partial}
\def\bea{\begin{eqnarray}}
\def\eea{\end{eqnarray}}
\def\ap{\alpha_+}
\def\am{\alpha_-}

\pagestyle{empty}
\rightline{UG-7/94}
\rightline{September 1994}
\rightline{hep-th/9409186}
\vspace{2truecm}
\centerline{\bf  On realizing the bosonic string as a noncritical $W_3$-string}
\vspace{2truecm}
\centerline{\bf E.~Bergshoeff, H.J.~Boonstra and M.~de Roo}
\vspace{.5truecm}
\centerline{Institute for Theoretical Physics}
\centerline{Nijenborgh 4, 9747 AG Groningen}
\centerline{The Netherlands}
\vspace{3truecm}
\centerline{ABSTRACT}
\vspace{.5truecm}

We discuss a realization of the bosonic string as a noncritical
$W_3$-string. The relevant noncritical $W_3$-string is
characterized by a Liouville sector which is restricted to a
(non-unitary) $(3,2)$ $W_3$ minimal model with central charge contribution
$c_l = - 2$. Furthermore, the matter sector of this $W_3$-string contains
$26$ free scalars which realize a critical bosonic string. The
BRST operator for this $W_3$-string
can be written as the sum of two, mutually anticommuting,
nilpotent BRST operators: $Q = Q_0 + Q_1$ in such a way that
the scalars which realize the bosonic string appear only in $Q_0$ while
the central charge contribution of the fields present in $Q_1$ equals zero.
We argue that, in the simplest case that the Liouville sector is given
by the identity operator only, the $Q_1$-cohomology is given by a
particular (non-unitary) $(3,2)$ Virasoro minimal model at $c=0$.

\vfill\eject
\pagestyle{plain}

\noindent{\bf 1. Introduction}
\vspace{.4cm}

It is well-known that the Virasoro algebra is the underlying symmetry
algebra of string theory. Extensions of the Virasoro algebra,
including generators of spin higher than two, are generically called
$W$-algebras. The simplest example is the so-called $W_3$-algebra
which contains an additional generator of spin three \cite{Za1}.
They can be used to construct new string theories, the
so-called $W$-strings \cite{Po1}.

A systematic investigation of multi-scalar matter
free-field realizations of the $W_3$-algebra
was undertaken in \cite{Ro1}. In particular, in \cite{Ro1} a general
mechanism was presented for generating realisations of the $W_3$-algebra,
given any realization of the Virasoro algebra, by adjoining an extra
matter scalar field, say $\phi_2$. All other scalars occur via their
energy-momentum tensor, say $T_\mu$. The
different scalars contribute to the matter central charge
$c_m = c_\mu + c_{\phi_2}$ as follows:
\begin{equation}
\label{eq:cc}
c_\mu = {1\over 4}c_m + {1\over 2}\,, \hskip 1.5truecm
c_{\phi_2} = {3\over 4}c_m - {1\over 2}\,.
\end{equation}
Note that the critical value $c_\mu = 26$ of the Virasoro algebra
does not lead to the critical value
\begin{equation}
c_m = 100
\end{equation}
of the $W_3$-algebra. Instead it leads to the anomalous value $c_m = 102$.
In \cite{Be1} it was pointed out that this excess of two units of central
charge could be compensated by passing from a ``critical'' to a
``non-critical'' $W_3$-string\footnote{Using the terminology
of \cite{Ber1}, a ``non-critical'' $W_3$-string
is characterized by the fact that there is a matter and Liouville sector
which {\sl separately} satisfy a $W_3$-algebra. If the Liouville sector is
absent we call the corresponding model a ``critical'' string theory.}
with a Liouville sector which is restricted to the
(non-unitary) $(3,2)$ model with central charge contribution
$c_l = - 2$.  The critical value of the central charge is now given
by
\begin{equation}
c_m + c_l = 100
\end{equation}
which is indeed satisfied by the values $c_m = 102$ and $c_l = - 2$.
Based on the above numerology, it was suggested in \cite{Be1} that
the above noncritical $W_3$-string should in some sense contain a critical
Virasoro string with $26$ free scalars.

After
the work of \cite{Be1} it was shown \cite{Po2} that, by going to a new
basis in the Hilbert space, the BRST operator of the $W_3$-algebra can be
written as the sum of two, mutually anticommuting, nilpotent
BRST operators
\begin{equation}
Q = Q_0 + Q_1\,.
\end{equation}
It was subsequently shown in \cite{Be2} that if one chooses for the Liouville
sector a $(p,q)\ W_3$
minimal model then the cohomology of the $Q_1$-operator
is given by a $(p,q)$ Virasoro minimal model. For
instance, the special case of a $(4,3)$ $W_3$
minimal model with central
charge $c_l=0$ leads to a $c_1 = 1/2$ Ising model in the $Q_1$-cohomology
\footnote{A similar result for the critical string was established earlier in,
e.g., \cite{Po3,Fr1,Hu2}.}.
In view of possible connections with the bosonic string we are particularly
interested in the case that the $Q_1$-cohomology is given by a (unitary)
$(3,2)$ Virasoro minimal model at $c_1=0$
which corresponds to just one state with conformal weight
$h_1=0$. The complete
$Q$-cohomology is then identical to that of a critical Virasoro
string. Unfortunately,
the arguments of \cite{Be2} are only valid for values of $(p,q)$ with
$(p,q) \ge (4,3)$. For these values there is a Kac-table characterizing
the finite set of primary fields of the model.

It is the purpose of this letter to investigate
the extrapolation
of the results of \cite{Be2} to values of $(p,q) $ with
$(p,q) < (4,3)$.
Having the connection with the bosonic string
in mind we will consider the particular case $(p,q) = (3,2)$.
We will argue that if one chooses for the Liouville sector
a (necessarily non-unitary) $W_3$ minimal model\footnote{
We will call a $(p,q)$ ($W_3$-)model {\it minimal}, if it is realized by
a finite number of\\
($W_3$-)primaries which form a closed
operator product algebra.
We infer the closure of the operator algebra from the
fusion rules \`a la BPZ \cite{BPZ}. This does not imply that the
non-unitary minimal models considered here necessarily
respect modular invariance. We thank Jan de Boer for a discussion
on this point.}
at $c_l = -2$ then
the $Q_1$-cohomology is given by a corrresponding non-unitary $(3,2)$
Virasoro minimal model at $c_1 = 0$. We will only give
explicit results for the simplest case that the $W_3$ minimal model is
given by the identity operator. The complete $Q$-cohomology contains,
among other states, all the states of a critical bosonic string.

\vspace{.4cm}

\noindent{\bf 2. The cohomology}
\vspace{.4cm}

In this section we will calculate the $Q_1$-cohomology at level zero
and level one.
In Table 1 we present the fields of the model with their background
 and central charges. Note that the matter sector contains $26$ free
scalars $X^\mu$ without a background charge.
 To facilitate the comparison with \cite{Be2}
 we have used a two-scalar realisation of the Liouville sector.

\vspace{.25cm}
\begin{center}
\renewcommand{\arraystretch}{1.5}
\begin{tabular}{|l||l|l|}
\hline
field  &\hfil background charge\hfil&\hfil central charge\hfil\\
\hline
$X^\mu$&$0$&$26$\\
$\phi_2$&$\tfrac{5}{2}$&$76$\\
\hline
$\sigma_1$&$-\tfrac{1}{6}i\sqrt{3}$&$0$\\
$\sigma_2$&$-\half i$&$-2$\\
\hline
$c,\ b$&\hfil&$-26$\\
$\gamma,\ \beta$&\hfil&$-74$\\
\hline
\end{tabular}
\renewcommand{\arraystretch}{1.0}
\end{center}
\vspace{.25truecm}

\noindent {\bf Table\ 1.} \ \ \ \ \ The fields of
the noncritical $W_3$-string. The scalar fields of the matter
sector ($X^\mu,\ \phi_2$) and of the Liouville sector
($\sigma_1, \sigma_2$) are given
with their background charge, and their contribution
to the central charge. The fields ($c,\ b$) are the spin-2,
($\gamma,\ \beta$) the spin-3 ghosts.
\vspace{.4truecm}

In \cite{Ber1,Be3} the BRST-operator for this system
was calculated.
It is given by $Q=\oint j$, with $j=j_0+j_1$ given by
\begin{eqnarray}
\label{J0}
 j_0 &=& c\,\{ T_M + T_L
     + T_{(\gamma,\beta)} +{1\over 2} T_{(c,b)} \} \,,
  \\
 j_1 &=&\gamma\,\big[\,
  {i\over 3\sqrt{6}}\,\{
   4(\p)^3 - 30\p\pp + 10\ppp \}
  \nonumber\\
  &&\qquad + i\,\{W_L- {2\over \sqrt{6}}\p T_L +{5\over 2\sqrt{6}}
  \partial T_L\}
   \nonumber\\
  \label{J1}
  &&\quad -i\sqrt 6\{ \p\partial\gamma\beta
+{5\over 6}\partial\beta\partial\gamma\}\big] \,.
\end{eqnarray}
$Q_0=\oint j_0$ and
 $Q_1=\oint j_1$ are seperately nilpotent, and therefore anticommute.
Note that the scalars  $X^\mu$, and the spin-2 ghosts are
 absent from $j_1$, and that the Liouville fields do not occur
 explicitly in (\ref{J1}), but only via the generators $T_L$ and $W_L$
which represent a standard two-scalar realization of the $W_3$-algebra.
In $j_0$ we find, besides $T_L$,
 the energy-momentum tensors of the matter and ghost sectors:
\begin{eqnarray}
  T_M&=& -\tfrac{1}{2}(\partial X^\mu)^2
         -\tfrac{1}{2}(\partial\phi_2)^2 + \tfrac{5}{2}
                  \partial^2\phi_2\,,\\
\label{Tgh2}
  T_{(c,b)} &=& -2 b \partial c - (\partial b) c \,,\\
\label{Tgh3}
  T_{(\gamma,\beta)}&=&  -3 \beta \partial \gamma -
      2 (\partial \beta) \gamma \,.
\end{eqnarray}

We first calculate all the states in the $Q_1$-cohomology
at level zero.
At level 0, the states of lowest ghost number $G=2$ are of the
 form\footnote{
At level $0$ it is enough to consider the states at lowest ghost number
$G=2$.
The states of higher ghost number $G=3$
can be obtained by acting with
picture-changing operators.}${}^{,}$\footnote{The level
 of a state in the $Q_1$-cohomology is defined by ${\rm level} \equiv h+3$,
 where $h$ is the weight of the fields in front of the exponential.
 In the case of the $Q$-cohomology (see below) we have
 ${\rm level} \equiv h+4$.}${}^{,}$\footnote{Our notation for the states is
 $V_{l,G}$ for the $Q_1$-cohomology and
 $U_{l,G}$ for the total cohomology. Here
 $l$ is the level and $G$ the ghost number.}
\begin{eqnarray}
\label{vac}
  V_{0,2}(p_2,s_1,s_2) &=& (\partial{\gamma})\gamma\, {\rm e}^{ip_2\phi_2+
             is_1\sigma_1+is_2\sigma_2} \,.
\end{eqnarray}
The condition
 $Q_1 V_{0,2}(p_2,s_1,s_2)=0$ determines the momenta of
 the three fields. The resulting
 cubic equation factorizes, and we obtain the
 following three solutions for $p_2$ and
 the corresponding conformal weight $h_1$ of $V$:
\begin{eqnarray}
\label{vacA}
  (A_0)&&p_2=i(s_2-2)\,,\\
       &&h_1=\tfrac{1}{2}s_1(s_1+1/\sqrt{3})\,,
 \nonumber \\
\label{vacB}
  (B_0)&&p_2=-\half i(s_2-\sqrt{3}\,s_1+5)\,,\\
       &&h_1=\tfrac{1}{8}
  (s_1+\sqrt{3}\,(1+s_2))(s_1+(1+3s_2)/\sqrt{3}) \,,
 \nonumber\\
\label{vacC}
  (C_0)&&p_2=-\half i(s_2+\sqrt{3}\,s_1+6)  \,,\\
       &&h_1=\tfrac{1}{8}
  (s_1-\sqrt{3}\,s_2)(s_1-(2+3s_2)/\sqrt{3})\,,\nonumber
\end{eqnarray}
where the Liouville momenta $s_1$ and $s_2$ are arbitrary.
We thus find a
two-parameter continuous spectrum of physical
 states at level zero.

It is not surprising that without any restriction on the Liouville sector
we find a continous spectrum. The same happens for the
non-critical $W_3$-strings studied in \cite{Be2}. In \cite{Be2}
a discrete spectrum
could be obtained by restricting the Liouville sector, by hand, to a
$W_3$ minimal model. In the present case we
cannot impose a similar restriction
since the Kac table corresponding to the $(p,q)=(3,2)$
Liouville sector is empty. However, in a first stage, it
is consistent to constrain the Liouville sector to the
completely degenerate representations, since these
representations form a closed operator algebra \cite{BPZ,FZ}.
Explicitly, the Liouville momenta can be put on the lattice
\bea
\label{LL}
s_1 &=& -{1\over 6} {\sqrt {3}}  (2r_2-3t_2)\,,\nonumber\\
s_2 &=& -{1\over 6}\bigl (2(2r_1-3t_1)+(2r_2-3t_2)\bigr )
\eea
for arbitrary non-negative integers $(r_1,r_2,t_1,t_2)$.
Substituting these Liouville momenta into the solutions
(\ref{vacA})-(\ref{vacC}), we find that the conformal weights
$h_1$ occurring in the $Q_1$-cohomology, correspond to the
completely degenerate representations of the $c=0$ Virasoro
algebra with conformal weights
\begin{equation}
h_{Vir} =\frac{1}{24} \biggl ((2r-3t-1)^2-1\biggr )
\end{equation}
for arbitrary non-negative integers $(r,t)$.

The $Q_1$-cohomology is now reduced from a
continuous to a discrete, albeit infinite, spectrum.
In \cite{Be2} the infinite spectrum could be reduced to a finite
spectrum by imposing further conditions involving the Kac table.
In the present case no such Kac table exists. However,
the Liouville sector can be restricted further
if there exist subsets of the completely degenerate representations
that form closed operator algebras.

At this point we may use the results of \cite{Bilal} where
fusion rules for $W_N$ algebra representations are given.
Using these fusion rules, it turns out that also in the $c_l=-2$ case
we can have closed fusion rules with a finite number of primaries.
We will continue our analysis for the simplest case
where we restrict the Liouville sector to
the unit operator with zero conformal and spin-3 weights.
Modulo translations, it has 6
representatives on the lattice
which are connected to each other by a Weyl transformation.
It turns out that four of these representatives
disappear in the cohomology with respect to the screening operators
of the $c_l=-2\ W_3$-algebra.
The remaining two representatives of the identity operator are given by
$(0,0;0,0)$ and
$(1,1;0,0)$. They are related to each other by
conjugation in the Liouville momenta. Ignoring the cohomology
with conjugate momenta
we can restrict ourselves
to the single point $(r_1,r_2;t_1,t_2) = (0,0;0,0)$ or
$(s_1,s_2) = (0,0)$ on the Liouville lattice (\ref{LL}).
This leads in (\ref{vacA})-(\ref{vacC}) to the weights
$h_1= (0,\tfrac{1}{8},0)$.

One can repeat the same analysis at level one.
Our results are summarized in Table 2. We obtain for ghost number
$G=1$ a new state with weight $\tfrac{5}{8}$. Other states in the $Q_1$
cohomology at this level are related to the $G=1$ states by
picture changing or conjugation\footnote{If excitations of the
Liouville fields $\sigma_{1,2}$ are included, we obtain at $G=2$
two states
with $h_1=1$, which are not descendants of the level 0, $h_1=0$ states.
However, these $h_1=1$ states disappear if
a Felder reduction of the free field
realisation of the Liouville minimal model is performed.
These states occur in the non-critical $W_3$-string
for a $(p,q)$ Liouville sector with arbitrary $p$ and $q$.}.

\vspace{.25cm}
\begin{center}
\renewcommand{\arraystretch}{1.5}
\begin{tabular}{|l||l|l|}
\hline
$G$  &\hfil level $0$\hfil&\hfil level $1$\hfil\\
\hline
$1$&$--$&$(0,\tfrac{5}{8})$\\
$2$&$(0,\tfrac{1}{8},0)$&$--$\\
\hline
\end{tabular}
\renewcommand{\arraystretch}{1.0}
\end{center}
\vspace{.25truecm}

\noindent {\bf Table\ 2.} \ \ \ \ \ The conformal weight $h_1$ of the
states in the $Q_1$-cohomology
at level $0$ and level $1$. States related to those shown
by picture-changing or conjugation of the momenta are not presented.
\vspace{.4truecm}

This concludes our analysis of the $Q_1$-cohomology at level zero and
level one. To obtain the total $Q$-cohomology we must dress the
states found in the $Q_1$-cohomology with spin-$2$ ghosts and $X^\mu$.
At level zero and lowest ghost number $G=3$ this gives the state
\begin{eqnarray}
\label{vaccgpp}
  U_{0,3}(p_X,p_2,s_1,s_2) &=&
     c\,(\partial{\gamma})\gamma\, {\rm e}^{ip_X\cdot X+ip_2\phi_2+
             is_1\sigma_1+is_2\sigma_2} \,.
\end{eqnarray}
The condition that $Q$ annihilates this state implies
 that $Q_0$ and $Q_1$ each annihilate this state\footnote{At
 higher levels this becomes more complicated.}.
 Thus in the $Q$-cohomology we
 find again the momenta (\ref{vacA}-\ref{vacC}), as well as the condition
 that $Q_0\,  U_{0,3}(p_X,p_2,s_1,s_2) =0$. This condition tells
 us that the total conformal weight of the physical state should vanish, and
 determines the value of $(p_X)^2$. We find:
\begin{equation}
\label{pX0}
  (p_X)^2= 2(1-h_1)\,,
\end{equation}
where $h_1$ is given in (\ref{vacA}-\ref{vacC}).

{}From (\ref{pX0}) we deduce
that for $h_1 = 0$ we have the (dressed) tachyon of the critical
Virasoro string with $(p_X)^2 = 2$.
It is not too difficult to see that for $h_1 = 0$ all the states of the bosonic
string occur. One just replaces the
$e^{ip_X\cdot X}$ part of (\ref{pX0}) by any bosonic string vertex operator
of the form $P(\partial X^\mu, \partial^2 X^\mu, \cdots)e^{ip_X\cdot X}$.
We thus conclude that any $h_1= 0$ solution in the $Q_1$-cohomology
leads to a bosonic string spectrum in the total $Q$-cohomology.
Alternatively, one can argue on more general grounds that the
bosonic string spectrum is contained in the total $Q$-cohomology.
Note that the total BRST operator $Q$ can be written as
$Q = Q_{\rm Vir} + Q_R$, where $Q_{\rm Vir}$ is the standard
BRST operator of the Virasoro string and $Q_R$ represents all other terms
in $Q$. Since $Q_R$ does not depend on $X^\mu$ and $b$ any bosonic string state
of the form $cV(X^\mu)$ is automatically $Q$-invariant. Sofar, our
explicit calculations indicate that such a state is never $Q$-trivial.

{}From the results given in Table 2 we conclude that
there are more states in the total $Q$-cohomology then the
bosonic string states. They correspond to
the $h_1 \ne 0$ solutions.
Since we have chosen in the Liouville sector a (non-unitary) $W_3$
minimal model at $c_l = -2$ consisting of the identity operator only
one would expect that the operators
in the $Q_1$-cohomology form
a corresponding (non-unitary) Virasoro minimal model at $c_1=0$.
In the next section we will present additional
arguments in support of this.

\vspace{.4cm}

\noindent{\bf 3. Other Methods}

\vspace{.4cm}

The explicit calculation of the $Q_1$-cohomology becomes increasingly
complicated at higher levels. It is therefore instructive to apply other
methods as well. In this section we will first apply a conjecture from
\cite{Be2} which is a formula summarizing the $Q_1$-cohomology in the
unitary case\footnote{
This conjecture is suggested by the analysis of \cite{Bo1}. A rigorous
proof is still lacking.}. Secondly, we
will also apply the fusion rules for Virasoro representations at
$c_1 = 0$.

\vspace{.4cm}

\noindent {\bf (1)}\ \ \ \ \ The formula of \cite{Be2}
gives a criterion for which values of the momenta $(p_2,s_1,s_2)$
a solution to the $Q_1$-cohomology can be expected.
This formula does not depend on the existence of a Liouville lattice and/or a
Kac table and can be applied in our case as well.
First, we define the polynomials $P_1, P_2^{\pm}$ and $P_3$ by\footnote{
These polynomials are specific to the case $(p,q) = (3,2)$.}
\begin{eqnarray}
P_1(n) &=& p_2 -is_2 + i (n+2)\,,\\
P_2^+(n) &=&p_2 + \tfrac{i}{2}s_2 - \tfrac{i}{2}s_1\sqrt 3 +
  \tfrac{i}{2}  (3n+5 )\,,\\
P_2^-(n) &=& p_2 + \tfrac{i}{2}s_2 - \tfrac{i}{2}s_1\sqrt 3 +
  \tfrac{i}{2} ( 2n+5 )\,,\\
P_3(n) &=& p_2 + \tfrac{i}{2}s_2 + \tfrac{i}{2}s_1\sqrt 3
+ \tfrac{3}{2}i(n+2)\,.
\end{eqnarray}
Restricting attention to Virasoro primaries the formula of \cite{Be2}
conjectures that the  $Q_1$-cohomology can be organized as
follows\footnote{
We give here a simplified version of the formula
in the sense that we do not give (i)
the level zero result which we have already given
in the previous section (it agrees with \cite{Be2}), (ii)
solutions which can be obtained by conjugation of the momenta and (iii)
a class of possible extra states which occur in the formula of \cite{Be2}
but which do not play a role in the present discussion.
}

\begin{itemize}

\item If integers $u_1, u_2 < 0$ exist such that (i) $P_1(u_1) =
P_2^+(u_2) = 0$, (ii) $P_2^-(u_1) = P_3(u_2) = 0$, or (iii) $P_1(u_1)
= P_3(u_2) = 0$ there are two states at level $u_1u_2$ and ghost
numbers $1,2$.

\end{itemize}

Applying this and restricting ourselves to the
identity operator on the Liouville lattice,
we reproduce all the states of Table 2.

As an example, consider the case that $P_1(u_1)=P_2^+(u_2)=0$.
This implies
\bea
p_2&=&-i(2+u_1-s_2)\nonumber\,,\\
s_1&=&{1\over\sqrt{3}}(1-2u_1 +3u_2 +3s_2)\,.
\eea
Clearly, a solution for $(s_1,s_2)=(0,0)$ can only exist if $2u_1 -3u_2=1$.
Substituting this in the formula for the conformal weight, we find
\bea
h_1=u_1u_2+\tfrac{1}{2}u_1(1-u_1)
\eea
for level $u_1u_2$. This shows that in this particular case all states
have integer conformal weight.
Taking the other cases into account as well, we find that there are only
states with weights $h_1 = n, \tfrac{1}{8} +n$ or
$\tfrac{5}{8} + n$ for non-negative integers $n$ in the $Q_1$-cohomology.

\vspace{.4cm}

\noindent {\bf (2)}\ \ \ \ \ For $(p,q) \ge (3,2)$ the fusion rules of the
Virasoro algebra give rise to the usual Kac tables.
For $(p,q) = (3,2)$ the Kac table only
consists of the identity, but it is not difficult to see that there
exist other choices of finite sets of primaries that are closed under
the BPZ fusion rules.
For instance, the fusion rules show that
$\{[0],[\tfrac{1}{8}],
[2],[\tfrac{1}{3}]\}$ form a closed set.
At level 1 we found in
the $Q_1$-cohomology a state with weight $\tfrac{5}{8}$.
If we include this primary in the previous minimal set
we obtain the following set of primaries that are closed under
the fusion rules:
$\{[0],[\tfrac{1}{8}],[2],[\tfrac{1}{3}],
[\tfrac{5}{8}],[1],[-\tfrac{1}{24}]\}$.
It seems that of these, the dimension $\tfrac{1}{3}$ and $-\tfrac{1}{24}$
operators do not occur in the $Q_1$-cohomology. The others do occur
in the $Q_1$-cohomology together with their descendants\footnote{
We assume that $[1]$ and $[2]$ are (null-)descendants of the
identity.}.
It is however not inconsistent that we don't have dimension
$\tfrac{1}{3}$
and $-\tfrac{1}{24}$ operators in the $Q_1$-cohomology. They may
decouple from the fusion rules.
In fact, using the relation between the conformal weight $h_1$, the level,
 and the momentum $p_2$ (for vanishing $s_1$ and $s_2$) one can see
 that the weight $\tfrac{1}{3}$ cannot occur in the OPE of two dimension
 $\tfrac{1}{8}$ operators. According to the fusion rules this would
 have been the place where dimension $\tfrac{1}{3}$ enters the algebra.
 With dimension $\tfrac{1}{3}$ absent, there is no longer any need for
 a  dimension $-\tfrac{1}{24}$ operator.

\vspace{0.4cm}

The two points above, together with our explicit calculations, indicate
 that with the Liouville sector restricted to
the identity operator, the $Q_1$-cohomology is given by a (non-unitary)
Virasoro minimal model at $c_1=0$ characterized by the
finite set of primaries $\{[0],[\tfrac{1}{8}],[\tfrac{5}{8}]\}$.

\vspace{.4cm}

\noindent{\bf 4. Conclusions}

\vspace{.4cm}

In this letter we have shown in which sense the bosonic string spectrum
occurs in the spectrum of a non-critical $W_3$-string with $c_l=-2$.
We have restricted our analysis to the simplest case that the
minimal model is given by the identity operator in the Liouville
sector. One could consider other minimal models as well.
The next to simplest model consists of three primaries
$\{ {\bf 1},\phi^+_k,\phi^-_k\}\,,\ k=1,2,3,...$, with spin-two weights
$\{0, k(2k-1),k(2k-1)\}$, respectively \cite{Bilal}. We expect that this
family of $W_3$ minimal models will lead to a corresponding family of
(non-unitary) $c=0$ Virasoro models in the $Q_1$-cohomology.
It would be interesting to see if choices are possible which respect
modular invariance.
%the precise relation is
%between an arbitrary $c_l=-2\ W_3$ minimal model in the Liouville sector and
%the corresponding $c_1=0$ Virasoro model in the $Q_1$-cohomology.

We expect that our results can be extended to the general case of a
critical $W_{n-1}$ string realized as a noncritical $W_N$-string.
The generalization of the $Q_1$-operator is a
BRST operator $Q_N^n$ corresponding to the subalgebra
$v_N^n \subset W_N$ consisting of generators of spin $s = \{n, n+1,
\cdots, N\}$\footnote{
We use the notation of \cite{Be2}.}.
We now restrict the Liouville sector
to a $(n,n-1)$ model of the $W_N$ algebra
with $(3 \le n \le N)$ and central charge
\begin{equation}
\label{eq:cl}
c_l = (N-1) \bigl \{1 - {N(N+1)\over n(n-1)}\bigr \}\,.
\end{equation}
In that case we find that
the central charge contributions of all fields present in
the $Q_N^n$ BRST operator add up to zero. This means that the
states in the $Q_N^n$-cohomology with zero weights lead to a
critical
$W_{n-1}$-string in the total cohomology.
The special case of $n=N=3$ corresponds to the
situation described in this letter.

Finally, we would like to mention recent work on the embedding of a
bosonic string into a ``critical'' $W_3$-string \cite{We1}. In this letter
we have described the embedding of a bosonic string into a
particular ``non-critical'' $W_3$-string. It would be
interesting to see whether there is any relation with the ``critical''
case.

\vspace{.4cm}

\noindent {\bf Acknowledgements}
\vspace{.4truecm}

We would like to thank Alexander Sevrin for a collaboration in an
early stage of this project and Jan de Boer for
enlightening  discussions.
The work of E.B.~has been made possible by a fellowship of the Royal
Netherlands Academy of Arts and Sciences (KNAW). The work of H.J.B.~was
performed as part of the research program of the ``Stichting voor
Fundamenteel Onderzoek der Materie'' (FOM).

\end{document}